\documentclass[12pt,preprint]{aastex}

\usepackage{epsfig}

\slugcomment{}

\shorttitle{Colors and ages of EROs-OGs}
\shortauthors{L\'opez-Corredoira}

\begin{document}


\title{Intrinsic colors and ages of 
extremely red elliptical galaxies at high redshift}


\author{Mart\'\i n L\'opez-Corredoira\altaffilmark{1,2}}
\affil{$^1$ Instituto de Astrof\'\i sica de Canarias, C/.V\'\i a L\'actea, s/n,
E-38200 La Laguna (S/C de Tenerife), Spain \\
$^2$ Departamento de Astrof\'\i sica, Universidad de La Laguna,
E-38205 La Laguna, Tenerife, Spain}

\email{martinlc@iac.es}



\begin{abstract}
In order to know the formation epoch of the oldest elliptical galaxies 
as a function of mass and observed redshift, a statistical analysis 
for 333 extremely red objects (EROs) classified as old galaxies (OGs) 
   at $0.8\le z\le 2.3$ is carried out.
Once we get $M_V$ and $(B-V)$ at rest for each galaxy, we 
   calculate the average variation of this intrinsic color with redshift
   and derive the average age through a synthesis model
   (the code for the calculation of the age has been made publicly available).
The average gradient of the $(B-V)$ color at rest of EROs/OGs 
   is 0.07--0.10 Gyr$^{-1}$ for a fixed luminosity.
   The stars in these extremely red elliptical galaxies were formed 
   when the Universe was $\sim 2$ Gyr old on average. We have not found
   a significant enough dependence on the observed redshift and stellar mass: 
   $\left(\frac{dt_{\rm formation}}{dt_{\rm observed}}=
   -0.46\pm 0.32\right)$, $\left(
   \frac{dt_{\rm formation}}{d\log _{10}M_*}=-0.81\pm 0.98\right)$ Gyr. 
   This fits a scenario in which the stellar formation of the objects
   that we denominate as EROs-OGs is more intense 
   at higher redshifts, at which the stellar populations of the
   most massive galaxies form earlier than
   or at the same time as less massive galaxies.
\end{abstract}





\section{Introduction}

It is usually sustained that galaxies at high redshift
are intrinsically bluer than at low redshift (e.g., Dickinson et al. 2003; 
Rudnick et al. 2006; Labb\'e et al. 2007), as would expected if
their populations were younger and with lower mass/luminosity ratios.
However, analysis of this intrinsic color evolution is not free
from caveats, due mainly to the difficulty in disentangling selection
effects. For instance, Dickinson et al. (2003, Fig. 2) showed
a clear lack of red objects between $2<z<3$ with 
$(m_{1700\AA }-m_B)_{AB,rest}>3$, while there were many of these 
red objects at $z<2$; so they consequently claimed that there is an evolution 
in color. However some red galaxies were probably missed in that analysis.
As a matter of fact, if we take the subsample of old elliptical galaxies 
without extinction ($f_{\rm old}\ge 0.95$) with $2<z<3$ 
from Miyazaki et al. (2003) ($N=12$ galaxies)
we find that all of them have $2.5<(m_{1700\AA }-m_B)_{rest}<4.5$ 
with an average of 3.6. Miyazaki et al.'s galaxies are much redder than
Dickinson et al.'s, which shows that the Dickinson et al. (2003)
sample does not contain the reddest galaxies.

On the question of the ages of galaxies and when were they formed,
there also some uncertainties.
There are  many observations at different redshifts of galaxies 
nearly as old as the Universe, although the density of those galaxies
at high redshift is significantly lower (Renzini 2006; Abraham et al. 2007). 
Early-type massive galaxies with early formation were found
at $z\sim 1$ (di Serego Alighieri et al. 2006; 2007),
$z=1-2$ (Spinrad et al. 1997; Daddi et al. 2005; Longhetti et al. 2005; 
Toft et al. 2005; Trujillo et al. 2006), 
$z=2-3$ (Daddi et al. 2005; Labb\'e et al. 2005; Cassata et al.
2008, Kriek et al. 2008, 2009), $z=3-4$ (Toft et
al. 2005; Chen \& Marzke 2004), $z=4-5$ (Chen \& Marzke 2004;
Rodighiero et al. 2007) and $z>5$ (Wiklind et al. 2008).
Most of this information is obtained from photometry, but 
there are also some spectra  massive elliptical galaxies at $z=1.4-2.2$ 
(Cimatti et al. 2008, Kriek et al. 2009) revealing them to be old galaxies.
Semi-analytical $\Lambda $CDM models 
(De Luc\'\i a et al. 2006) claim that the formation 
of low mass galaxies is first to give way to mergers
of very massive galaxies, which is apparently at odds
with observation of these massive galaxies at high redshift, 
even if they were formed through dry mergers in a downsizing scenario.
Ferreras et al. (2009) also found that very massive galaxies
do not have significant evolution at $z<1.2$.
However, Schiavon et al. (2006), who have taken spectra of 
red galaxies ($U-B>0.25$) at $z\sim 0.9$, 
derived their age  to be
on average 1.2 Gyr, which means that some galaxies were formed at
lower redshifts. Arnouts et al. (2007) also show evidence for 
a major build up of the red sequence between $z=2$ and $z=1$.

In this paper, we pay further attention to the determination of
intrinsic color and age variation for different redshifts in a 
statistical way for two different samples of very red elliptical galaxies
in order to constraint their formation epoch. 

\section{Data}

We select galaxies classified as 
Extremely Red Objects (EROs) within 
the redshift range $0.8\le z\le 2.3$
and, within this group, those classified as Old Galaxies (OGs) 
with negligible intrinsic extinction; that is, 
passively evolving populations of elliptical
galaxies. The different methods of selecting OGs are quite consistent with
each other (Fang et al. 2009). 
Galaxies will be selected with available 
fluxes in the three near infrared filters (JHK), plus
at least two filters in the optical for which the flux signal/noise  is 
greater than 3.
We use two sources of publicly available data:

\begin{enumerate}

\item ECDFS catalog (Taylor et al. 2009a), which gives photometry 
in ten filters: U, U$_{38}$, B, V, R, I, z', J, H, K$_s$
from ISAAC(VLT)+Hubble data. For the selection of EROs-OGs,
we adopt the color criterion $(i_{775}-K)_{AB}>2.42$ (EROs), 
$(J-K)_{AB}<0.20(i_{775}-K)_{AB}+0.39$ (Fang et al. 2009). 
We do not have the magnitude at $i_{775}$ 
but we get this with the corresponding color correction using
the adjacent filters. A total of 276 galaxies.

\item Miyazaki et al. (2003) give photometry 
in eight filters: B, V, R, i', z', J, H, K$_s$ from Subaru/XMM-Newton+UH2.2m. 
Their sample of EROs-OGs was selected using $(R-K)_{AB}>3.35$ (EROs),
and within these sources, by means of spectrum fitting,
as OGs without extinction with a fraction of old population
$f_{old}\le 0.95$. A total of 57 galaxies.

%


\end{enumerate}

Estimating roughly the average intrinsic color/age 
of EROs/OGs is our aim here. 
The term ``ERO'' reflects the observed characteristic  
color of a galaxy, not its intrinsic properties. For this reason, and
because of using magnitude-limited samples, they have different 
ranges of stellar masses, M/L ratios and intrinsic colors at different 
redshifts. It is not a homogeneous sample of galaxies with the same
intrinsic characteristics at all redshifts; there are biases.
Nonetheless, throughout the paper we shall separate the dependence on redshift
from the dependence on luminosity/mass.

\section{Color $(B-V)$ at rest}
\label{.colors}

We take AB apparent magnitudes, corrected for Galactic extinction (although
negligible), for different wavelengths, $m_{AB}(\lambda _i)$, $(i=1,...,N_f)$,
for $5\le N_f\le 10$, with the corresponding error bars 
(in our case in optical and near infrared). As has already been said, we consider only 
the points with a flux signal/noise above 3. We will use
data with available redshifts, most of them photometric.
The average systematic error of the photometric redshifts is 
$\frac{\Delta z}{(1+z)}\sim -0.025$ for ECDFS (Taylor et al. 2009a, \S 7.3), 
which is small, so we do not take it into account here; 
similarly for the Miyazaki sample.
There is a statistical error for each photometric
redshift, but we expect that these uncertainties will nearly cancel in
the statistical analysis.

With this, we calculate the rest luminosities  in two filters.
This is done through  spectral energy distribution (SED) fitting using
templates of galaxies with the software Interrest v2.0 (Taylor et al. 2009a).
The calculations for the sample ECDFS have already been carried out by Taylor
et al. (2009a). The calculation for Miyazaki et al. (2003) was carried out
by us with the Interrest v2.0 software, by changing to the Subaru filters.
In this paper, we do the calculations only with a pair of filters
at rest (Johnson B and V), 
applying the correction to convert AB into the Vega calibration:
$(B-V)=(B-V)_{AB}+0.119$ (Frei \& Gunn 1994).
The $(U-B)_{\rm rest}$ color is not used in this paper 
because it is more sensitive to
redshift uncertainties and uncertainties in the emission-line
corrections (Rudnick et al. 2006). Moreover, the $(U-B)$ color depends
more strongly than $(B-V)$ on metallicity for a given age; 
it is also more sensitive to $\alpha $-enhancement (Cassisi et al. 2004).

There are uncertainties in the rest color, due to the error bars
of the apparent magnitudes, the deviation of the
assumed shape from the true SED (spectral features which move through
the filter bands), errors in the photometric redshifts, etc.
In any case, we do not expect important systematic errors, and the statistical
errors can be reduced when we calculate the average for bins with a large
number of galaxies.

\begin{figure}
\begin{center}
\vspace{1cm}
\mbox{\epsfig{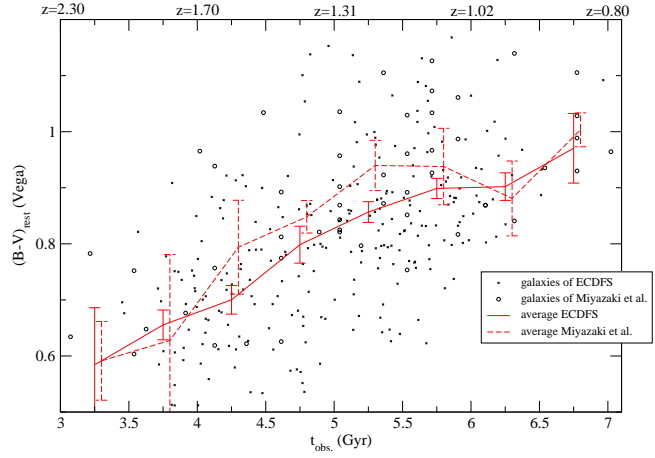}}
\end{center}
\caption{$(B-V)_{\rm rest}$ (Vega calibration) colors observed
at different ages $t_{\rm obs.}$.}
\label{Fig:colors}
\end{figure}

In Fig. \ref{Fig:colors}, we give the
results of the average colors
as a function of the age of the Universe when the galaxy is observed,
\begin{equation}
t_{obs.}(z)=\frac{1}{H_0}\int _\infty ^z dx \frac{-1}{1+x}
\frac{1}{\sqrt{\Omega _m(1+x)^3+\Omega _\Lambda }}
\label{ageuniv}
,\end{equation}
with $H_0=73$ km/s/Mpc, $\Omega _m=0.24$, $\Omega _\Lambda =0.76$.
Both samples give approximately the same results; thus the selection of
OGs among EROs is shown to be quite consistent with both independent methods.
There is a significant average gradient in color of $0.124\pm 0.014$ and
$0.118\pm 0.021$ Gyr$^{-1}$ respectively for samples ECDFS and Miyazaki et al.

Note that the galaxy gap in the lower right corner 
of Fig. \ref{Fig:colors} is at least partially an artifact of the sample
intrinsic color bias as a function of the redshift. This is expected
because at the lowest redshift, the $(i-K)$ and $(R-K)$ limits used will
pick out only the very oldest and reddest galaxies, whereas, at higher
redshifts, younger and bluer galaxies will be included. Note even that
at the very lowest redshifts in Fig. \ref{Fig:colors} the much smaller
Miyazaki sample dominates; this is because the ECDFS criterion is a bit
more strigent, so it eliminates even the oldest passive galaxies at $z\sim 0.8$.
One must therefore interpret Fig. \ref{Fig:colors} as the
intrinsic color as a function of redshift of the objects selected as EROs/OGs, 
not of a general characterization for elliptical galaxies.

If we do a bi-linear fit of the colors as a function of 
two independent variables $t_{obs.}$ and $M_{V,rest}$,
we get: 
\begin{equation}
(B-V)_{rest}=a_1+a_2[t_{obs.}({\rm Gyr})-5]
+a_3(M_{V,rest}+22)
\label{bmv2fit}
,\ \end{equation}
with $a_1=0.756\pm 0.014$, $a_2=0.084\pm 0.015$, $a_3=0.065\pm 0.013$ for
ECDFS; $a_1=0.814\pm 0.023$, $a_2=0.087\pm 0.027$, 
$a_3=0.048\pm 0.028$ for Miyazaki.
This bi-linear fitting allows us to separate the evolution 
from the biases in absolute magnitudes; so the second term (0.084 or
0.087) gives us the average evolution in color for a fixed luminosity.
Galaxies are redder for lower observed redshift
and for lower luminosity. The first fact indicates older galaxies at
lower redshift, and the second fact is probably  related to a higher
luminosity for younger populations.
Other authors, for instance Labb\'e et al. (2007) have found 
however that the most luminous galaxies have redder colors,
with a slope of $\frac{d(U-V)}{dM_V}= -0.09\pm 0.01$ (Labb\'e et al. 2007).
Our guess is that we do not find the same luminosity dependence because
we have preselected massive galaxies with the constraint that they be EROs/OGs, and
within a different range of redshifts.
In Taylor et al. (2009b, Fig. 4), 
we see that $\frac{d(u-r)}{dM_r}$ is negative for $z<1.25$; 
however,  the trend changes for $z>1.25$ and this variation
of average color with absolute magnitude is null or even slightly positive.

\section{Age estimation for early-type galaxies}

{\it NOTE: A FORTRAN code
to carry out the calculations explained in this section is
available at http://www.iac.es/galeria/martinlc/codes.html}

As has been said, we have selected old massive elliptical galaxies 
with negligible internal extinction, i.e. without gas and dust. 
We can therefore connect the color and luminosity
in V-rest of the galaxy with the average age of the stellar population
and its metallicity. 
There may be some wrong identification of OGs among our galaxies selected
with the color method for the ECDFS sample (Miyazaki et al. 2003 and 
Fang et al. 2009 estimate it to be $\sim 25$\%),
and consequently there may be some case of dusty galaxies among our
ECDFS sample, but the statistical comparison with the SED fitting method of the
Miyazaki et al. sample in Fig. \ref{Fig:colors} shows that there 
are neither significant differences nor systematic effects with 
redshift. Only perhaps in the range $1.2<z<1.6$
might there be some small difference, where the starburst contamination
might be higher (Fang et al. 2009).

In order to estimate the average age corresponding to our galaxies, we 
use a synthesis model: Vazdekis et al. (1996; hereafter V96); 
see also Vazdekis (1999). There is an age--metallicity degeneracy, but this can
be broken approximately with the use of the mass--metallicity correlation.
Another way to break the degeneracy would be by using two colors (Li \& Han 
2007, and references therein), but we have only one reliable color at 
rest and we do not have rest near-infrared  colors as necessary in Li 
\& Han (2007).

We must also bear in mind that
synthesis models return the mean value of a distribution, and a
perfect fit to observational data to infer the age is only correct on
average, since the individual cases may present some dispersion
with respect to the average (Cervi\~no \& Luridiana 2006).
For this reason, and because of the large errors in the color of each
galaxy, we do not calculate the age of each galaxy separately
but the average age of each bin of galaxies (11 in our case) with the
same redshift. 
 
\begin{figure}
\begin{center}
\vspace{1cm}
\mbox{\epsfig{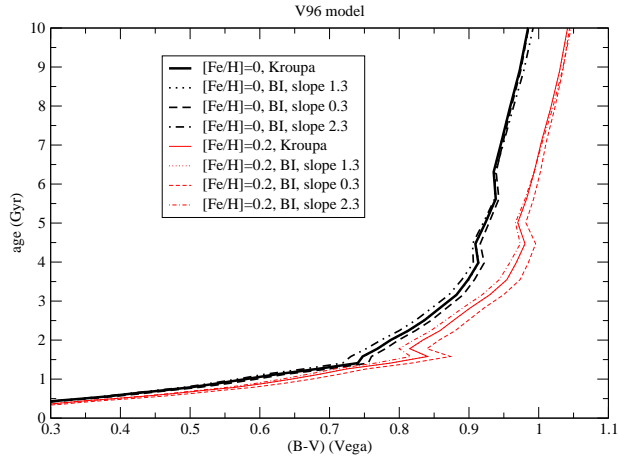}}
\end{center}
\caption{Variation of age with $(B-V)$ color in the V96 stellar population
synthesis model.}
\label{Fig:bmv_age}
\end{figure}

For each bin of galaxies, the steps are as follows:

\begin{enumerate}

\item We assume zero metallicity and derive 
the age ($t_1$) of the galaxy which is given by the V96 
model for the given $(B-V)_{\rm rest}$ of the galaxy.

\item Given the age $t_1$ and the zero metallicity, we derive with
the V96 model the stellar mass-to-light ratio in the V  filter [$(M_*/L_V)_1$].

\item Since we know the luminosity at rest in the V filter ($L_V$), 
we can derive the stellar mass of the galaxy:
\begin{equation}
M_{*,1}=(M_*/L_V)_1L_V
\label{mlratio}
.\end{equation}


\item Given the stellar mass of the galaxy, we estimate the metallicity 
$[Fe/H]_1$. We use the correlations of metallicity and $\alpha $-enhancement 
given by Thomas et al. (2005). The average relation is:

\begin{equation}
[Fe/H]=0.066\log _{10}\left(\frac{M_*}{1.1\times 10^{11}\ M_\odot}
\right)\pm 0.11
\label{metal}
,\end{equation}
with and uncertainty of $\delta ([Fe/H])\approx 0.11$ including the scatter
of the correlations and the variations with the environment 
(low or high densities) of the galaxies.
The correlation of mass (or velocity dispersion) with metallicity
is also observed in Yamada et al. (2007). There is also a dependence
on age, but it is only important for low mass galaxies of
velocity dispersion less than 100 km/s (Yamada et al. 2007), which
is not the case of our galaxies.
It is assumed that the relationship of mass and metallicity does
not evolve with redshift in a passive evolution
(di Serego Alighieri et al. 2006).
A metallicity evolution in galaxies of the same mass
is not observed (Cimatti et al. 2008) so, provided that mass does 
not correlate with the age, the mass--metallicity relationship should 
not change too much at high redshift.
As  is observed in Fig. \ref{Fig:bmv_age}, the accuracy in the metallicity
determination mainly  affects the reddest (oldest) galaxies. For the youngest
galaxies ($<3$ Gyr) the errors in metallicity are not very important, so a
possible evolution in the relationship of Eq. (\ref{metal}) at high 
redshift would not affect the results.

\item We derive again the age $t_2$ with the color $(B-V)_{\rm rest}$ 
and metallicity $[Fe/H]_1$.

\item We repeat steps 2--5, which gives an age $t_3$. Since $t_3\approx t_2$,
we do not need to do further iterations and we have got the convergence
of the metallicity, mass and age with only three iterations. If $t_3$ were
significantly different from $t_2$, we would continue to iterate to get
$t_4$, $t_5$,... until the convergence is obtained.

\end{enumerate}

There is some dependence with the IMF (initial mass function) slope. 
In Fig. \ref{Fig:bmv_age},
we plot age vs. color for metallicities $[Fe/H]=0$ and $[Fe/H]=0.2$ and
different slopes in the bimodal IMF, as defined in V96, 
or Kroupa IMF using V96 model. 
Slope 1.3 in the bimodal IMF is the standard value
and is nearly coincident with the Kroupa case. Within variations
of $\pm 1$ of the slope the variations of the age are fitted from
Fig. \ref{Fig:bmv_age} for null metallicity by:
\[
\frac{(\Delta t)^+}{t}=-0.071+0.346(B-V)-0.257(B-V)^2
,\]\begin{equation}
\frac{(\Delta t)^-}{t}=0.123-0.299(B-V)+0.243(B-V)^2
\label{err_IMF}
.\end{equation}
These variations produce some error in the age determination.
However, this error is smaller than that produced by the uncertainty
in the color or the metallicity. In any case, they are taken into account.

\begin{figure}
\begin{center}
\vspace{1cm}
\mbox{\epsfig{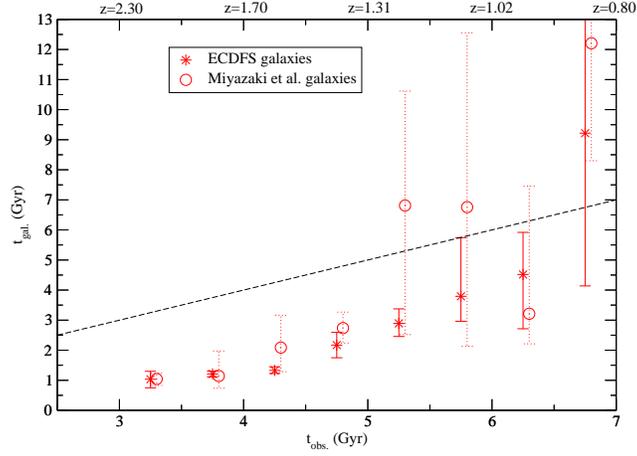}}
\end{center}
\caption{Average age of the stellar populations of EROs which are
passively evolving early-type galaxies.
The dashed line stands for the limiting maximum $t_{gal}=t_{\rm obs.}$ 
within the standard cosmology.}
\label{Fig:ages}
\end{figure}

The age of the early-type galaxies is plotted
in Fig. \ref{Fig:ages}. 
The vertical bars include the errors due to uncertainties in the
colors, the uncertainty of 0.11
in the mass-metallicity relationship of Eq. (\ref{metal}) and the variations
due to the IMF slope change within a range of $\pm 1$ given by Eq.
(\ref{err_IMF}).
There is no zero-point calibration problem except perhaps for 
the last bin because this affects mainly galaxies older
than 5 Gyr (Vazdekis et al. 2001). 
We must also bear in mind that we are neglecting the systematic errors
in the photometric redshifts; were they non-negligible, we would have
extra systematic errors in the calculated ages.

Figure \ref{Fig:ages} represents the average age
of the given sample among the ERO/OGs, with all the
selection effects associated with each redshift.
If we separate the dependence on mass from its evolution, 
subdividing each redshift bin into sub-bins with 
different luminosity ($\Delta M_V=1$), 
and we do over them a bilinear fit weighted with the square inverse of 
relative errors, both for average color and average age, we get 

\begin{equation}
(B-V)_{rest}=b_1+b_2(t_{\rm obs.}-5)
+b_3\log_{10}(M_*)
\ ,\end{equation}
\begin{equation}
t_{\rm gal.}=c_1+c_2(t_{\rm obs.}-5)
+c_3\log_{10}(M_*)
\ ,\end{equation}
with $t_{\rm gal.}$ and $t_{\rm form.}$ in units of Gyr, $M_*$ in units
of $10^{11}\ {\rm M_\odot}$;
$b_1=0.779\pm 0.025$, $b_2=0.106\pm 0.014$, $b_3=-0.042\pm 0.049$,
$c_1=2.22\pm 0.54$, $c_2=1.15\pm 0.32$, 
$c_3=-0.4\pm 1.0$ for ECDFS; and 
$b_1=0.845\pm 0.020$, $b_2=0.085\pm 0.022$, $b_3=0.0027\pm 0.0020$,
$c_1=5.06\pm 0.96$, $c_2=2.71\pm 0.78$, 
$c_3=4.0\pm 2.1$ 
for Miyazaki. Average stellar masses in the bins 
range between $5\times 10^9$ $M_\odot $ and $3\times 10^{11}$ $M_\odot $.
Kriek et al. (2008) found in the same range of redshifts 
a $\frac{d(U-B)_{\rm rest}}{dt_{obs.}}=+0.024$
Gyr$^{-1}$  for the red sequence at a given mass of 
$2\times 10^{11}$ M$_\odot $, equivalent (with $\frac{d(B-V)}{d(U-B)}\approx
0.8$; V96 at [Fe/H]=0, $(U-B)=0.25-0.37$) 
to $b_2=\frac{d(B-V)}{dt_{obs.}}\approx
+0.03$ Gyr$^{-1}$, a smaller color
evolution but for larger masses than our sample.

\section{Discussion on their formation epoch}
\label{.form}

\begin{figure}
\begin{center}
\vspace{1cm}
\mbox{\epsfig{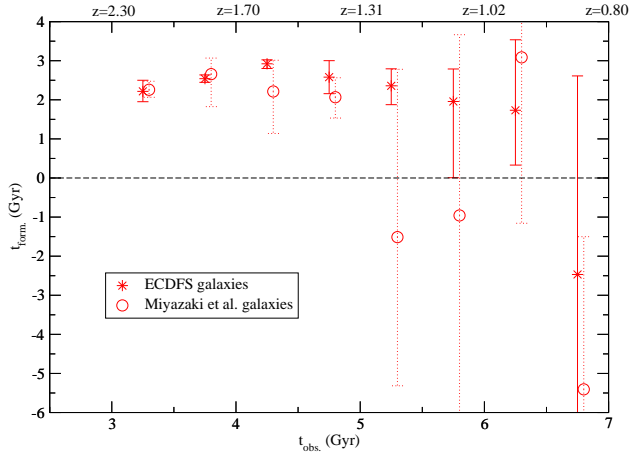}}
\end{center}
\caption{Average age of the Universe at which the stellar populations of the
galaxies observed at age $t_{\rm obs.}$ have been formed.}
\label{Fig:ages_form}
\end{figure}

The average epoch of star formation (the epoch of formation of the first 
stars, might be lower) is 
$t_{\rm form.}=t_{\rm obs.}-t_{\rm gal.}$, 
shown in Fig. \ref{Fig:ages_form}. Separating the evolution from the
mass dependence, 
\begin{equation}
t_{\rm form.}=d_1+d_2(t_{\rm obs.}-5)+
d_3\log_{10}(M_*)
,\end{equation}
with $d_1=1.94\pm 0.51$, $d_2=-0.46\pm 0.32$, $d_3=-0.81\pm 0.98$
on average for the ECDFS+Miyazaki samples.
The observed EROs/OGs are all formed within a narrow
range of epochs, when the Universe was less than
4 Gyr old ($z>1.7$). The present fit is compatible with all EROs/OGs formed
at the same time at, on average, $t_{\rm form.}=2.0\pm 0.3$ Gyr ($z\sim 3-4$).
Again, we remind the reader that the criterion to select EROs/OGs
is more restrictive at lower redshift, picking out only very old galaxies,
while at higher redshift the range of allowed ages is wider.
We might therefore expect that the oldest EROs at high $z$ (low 
$t_{\rm obs.}$) will have an even lower formation age. 
The age $t_{\rm form.}=2.0$ Gyr is a conservative lower limit representing the
average sample; there must be some EROs/OGs formed beforehand.
Given that $\frac{dt_{\rm form.}}{dt_{\rm obs.}}=-0.46\pm 0.32<<+1$ for a
given stellar mass, this means that the galaxies of the present sample are
not formed continuously with the same rate, but more intensely at higher 
redshifts.

The stellar populations of most massive galaxies are not 
formed much later than the less massive ones ($d_3\le 0$).
This agrees the results mentioned in the introduction that
very massive evolved galaxies detected at
redshifts 1.5--6 were formed in the very early Universe (Daddi et al. 2005:
Chen \& Marzke 2004; Rodighiero et al. 2007; Wiklind et al. 2008).
This might appear in contradiction
with the result of \S \ref{.colors} that galaxies are redder for lower 
luminosities, but it does not. As said, the mass--luminosity ratio does 
not remain constant giving higher luminosity for younger objects so it is
not contradictory that older/redder objects correspond to lower luminosities
and higher masses. It is in fact observed if we compare Figs. 4 and 5 
of Taylor et al. (2009b) for $z>1.25$: clearly, a strong dependence 
on stellar mass does not mean a strong dependence on luminosity.

\

{\bf Acknowledgments:}
I thank the anonymous referee for helpful comments in
correcting and improving this paper, to A. Vazdekis (IAC) for helpful suggestions on
the use of his V96 model and comments on a draft of this paper,
and to T. J. Mahoney (IAC) for proof-reading this paper.
The author was supported by the {\it Ram\'on y Cajal} Programme
of the Spanish Ministry of Science.


\begin{thebibliography}{99}

\bibitem{} Abraham, R. G., Nair, P., McCarthy, P. J., et al.
2007, ApJ 669, 184

\bibitem{} Arnouts, S., Walcher, C. J., Le F\`evre, O., et al. 2007,
A\&A 476, 137

\bibitem{} Cassata, P., Cimatti, A., Kurk, J., et al. 2008,
A\&A 483, L39

\bibitem{} Cassisi, S., Salaris, M., Castelli, F., \& Pietrinferni, A.
2004, ApJ 616, 498

\bibitem{} Cervi\~no, M., \& Luridiana, V. 2006, A\&A 451, 475

\bibitem{} Chen, H.-W., \& Marzke, R. O. 2004, ApJ 615, 603 

\bibitem{} Cimatti, A., Cassata, P., Pozzetti, L., et al.,
2008, A\&A, 482, 21

\bibitem{} Daddi, E., Renzini, A., Pirzkal, N., et al. 2005, ApJ
626, 680 

\bibitem{} De Luc\'\i a, G., Springel, V., White, S. D. M., 
Croton, D., \& Kauffmann, G. 2006, MNRAS 366, 499

\bibitem{} di Serego Alighieri, S., Bressan, A., Pozzetti, L.
2007, in: From Stars to Galaxies: Building the Pieces to Build Up 
the Universe (ASP Conf. Ser. 374), A. Vallenari, R. Tantalo, 
L. Portinari, A. Moretti, ASP, S. Francisco, Eds, p.\ 449

\bibitem{} di Serego Alighieri, S., Lanzoni, B., Jorgensen, I.
2006, ApJ 647, L99. Erratum at: 2006, ApJ 652, L145

\bibitem{} Dickinson, M., Papovich, C., Ferguson, H. C., 
Budav\'ari, T. 2003, ApJ, 587, 25

\bibitem{} Fang, G.-W., Kong, X., \& Wang, M. 2009, Res. Astron. 
Astrophys. 9, 59

\bibitem{} Ferreras, I., Lisker, T., Pasquali, A., Khochfar, S.,
\& Kaviraj, S. 2009, MNRAS 396, 1573

\bibitem{} Frei, Z., \& Gunn, J. E. 1994, AJ 108, 1476

\bibitem{}
Kriek, M., van der Wel, A., van Dokkum, P. G., Franx, M., \& Illingworth, G. 
D. 2008, ApJ 682, 896

\bibitem{} 
Kriek, M., van Dokkum, P. G., Labbe, I., Franx, M., 
Illingworth, G. D., Marchesini, D., \& Quadri, R. F. 2009, ApJ 700, 221

\bibitem{} Labb\'e, I., Franx, M., Rudnick, G., et al. 2007, ApJ 665, 944

\bibitem{} Labb\'e, I., Huang, J., Franx, M., et al. 2005, ApJ 624, L81

\bibitem{} Li, Z., \& Han, Z. 2008, MNRAS 385, 1270

\bibitem{} Longhetti, M., Saracco, P., Severgnini, P., et al.
2005, MNRAS, 361, 897

\bibitem{} Miyazaki, M., Shimasaku, K., Kodama, T., et al. 2003,
PASJ 55, 1079

\bibitem{} Renzini, A. 2006, ARA\&A 44, 141

\bibitem{} Rodighiero, G., Cimatti, A., Franceschini, A.,
Brusa, M., Fritz, J., \& Bolzonella, M. 2007, A\&A 470, 21

\bibitem{} Rudnick, G., Labb\'e, I., F\"orster Schreiber, N. M.,
et al. 2006, ApJ 650, 624

\bibitem{} Schiavon, R. P., Faber, S. M., Konidaris, N., et al.
2006, ApJ 651, L93

\bibitem{} Spinrad, H., Dey, A., Stern, D., Dunlop, J.,
Peacock, J., Jim\'enez, R., \& Windhorst, R. 1997, ApJ 484, 581

\bibitem{} Taylor, E. N., Franx, M., van Dokkum, P. G., et al.
2009a, ApJS 183, 295

\bibitem{} Taylor, E. N., Franx, M., van Dokkum, P. G., et al.
2009b, ApJ 692, 1

\bibitem{} Thomas, D., Maraston, C., Bender, R., \& Mendes de Oliveira, C.
2005, ApJ 621, 673

\bibitem{} Toft, S., van Dokkum, P., Franx, M., Thompson, R. I., 
Illingworth, G. D., Bouwens, R. J., \& Kriek, M. 2005, ApJ 624, L9

\bibitem{} Trujillo, I., Feulner, G., Goranova, Y., et al. 2006,
MNRAS 373, L36

\bibitem{} Vazdekis, A. 1999, ApJ 513, 224

\bibitem{} Vazdekis, A., Casuso, E., Peletier, R. F., \& Beckman, 
J. E. 1996, ApJS 106, 307 (V96)

\bibitem{} Vazdekis, A., Salaris, M., Arimoto, N., \& Rose, J. A.
2001, ApJ 549, 274

\bibitem{} Wiklind, T., Dickinson, M., Ferguson, H. C.,
Giavalisco, M., Mobasher, B., Grogin, N. A., \& Panagia, N.
2008, ApJ, 686, 781

\bibitem{} Yamada, Y., Arimoto, N., Vazdekis, A., \& Peletier,
R. F. 2006, ApJ, 637, 200

\end{thebibliography}
\end{document}